\documentstyle[11pt,newpasptello,twoside,epsf]{article}
\def\1{\'\i}                    
\def\3{\c c}                   
\def\4{\"u\^e}                           
\def\ao{\~ao\ }                             
                                                        
\def\th{\thinspace}

\def\sy{\scriptscriptstyle}                                    
\def\lsim{\,\raise 0.25ex\hbox{$<$}\hskip -0.7em \raise -0.45ex                    
\hbox{$\scriptstyle\sim$}\hskip 0.2em\,}                   
\def\gsim{\,\raise 0.25ex\hbox{$>$}\hskip -0.7em \raise -0.45ex                  
\hbox{$\scriptstyle\sim$}\hskip 0.2em\,}              
\def\llsim{\raise 0.26ex\hbox{$\scriptscriptstyle <$}\hskip -0.5em \raise -         
0.45ex\hbox{$\scriptscriptstyle\sim$}\hskip 0.2em}                   
\def\ggsim{\raise 0.26ex\hbox{$\scriptscriptstyle >$}\hskip -0.5em \raise -         
0.45ex\hbox{$\scriptscriptstyle\sim$}\hskip 0.2em}                 
\def\asolm{\lower 1.1em\hbox{\hskip -1.5em$\sy \Omega_M$\hskip 1ex}}               
\def\asol{\lower 1.1em\hbox{\hskip -1.5em$\sy \Delta\Omega$\hskip 1ex}}                  
\def\a4pi{\lower 1.1em\hbox{\hskip -1.4em$\sy 4\pi$}\hskip 1ex} 
\def\afence{\lower 1.2em\hbox{\hskip -1.7em$\sy \Omega_f$\hskip 1ex}}                                     
\newbox\strutbox                   
\setbox\strutbox=\hbox{\vrule height11pt depth3.5pt width0pt}

\def\mag{,\hskip -0.37em \raise 1ex\hbox{$\scriptstyle m$}}                   
\def\beq{\begin{equation}}                  
\def\eeq{\end{equation}}                  
\def\beqy{\begin{eqnarray}}                  
\def\eeqy{\end{eqnarray}}

\def\pin#1{\put(10,4){\circle{12}}\raise 0.2ex\hbox{\hskip        
1.45ex{\small{#1}}\hskip 1.45ex}}                       
\def\pn#1{\put(10,4){\circle{14}}\raise 0.2ex\hbox{\hskip        
0.82ex{\small{#1}}\hskip 1ex}}       
\def\fa{408\thinspace{MHz}}         
\def\fb{1465\thinspace{MHz}}  
\def\fc{$2.3$\th{GHz}} 
      
\def\fe{10\th{GHz}}
\def\ord#1{\raise 0.7ex\hbox{\small\underbar{#1}}}        
\def\ordd#1{\raise 0.8ex\hbox{\tiny\underbar{#1}}} 

\begin{document}
\title{A New Radio Continuum Survey of the Sky at 1465 MHz between declinations 
$-52\deg$ and $+68\deg$}
 \author{C. Tello, T. Villela}
\affil{INPE, CP 515 S\ao Jos\'e dos Campos, SP, 12201-970, Brazil}
\author{G.F. Smoot}
\affil{LBNL, MS 50-205, Berkeley, CA, 94720, USA}
\author{S. Torres}
\affil{Centro Internacional de F\1sica, Bogot\'a, Colombia}
\author{M. Bersanelli}
\affil{Dpto. di Fisica, Univ. di Milano, Via Bassini 15,  20133 Milano, Italy}

\begin{abstract}
We have mapped the total sky brightness at \fb\ in two adjacent $60\deg$ 
declination bands with the portable 5.5-m parabolic reflector of the Galactic 
Emission Mapping (GEM) project, an on-going international collaboration to 
survey the radio continuum of the sky in decimeter and centimeter wavelengths. 
The observations were conducted from two locations, one in the USA and the 
other in Brazil, using a novel instrumental approach to overcome the 
well-known shortcomings of survey experiments. Our strategy consists of a 1-rpm rotating 
dish to circularly scan the sky at $30\deg$ from zenith. The dish uses a rim-halo 
to re-direct the spillover sidelobes of its backfire helical feed toward the sky and 
the entire assembly has been enclosed inside a wire mesh ground shield in order 
to minimize and level out the contamination from the ground. The diffraction 
characteristics of this set-up have been succesfully modelled and undesired 
systematic striping across the observed bands has been carefully removed by a 
baseline propagation method which exploits the time-forward and time-backward 
intersections of the circular scans. The map displays nearly 300~hours of our best 
quality data taken with a HPBW of $5.4\deg$ at a sensitivity of 20 mK. 
\end{abstract}

\section{Introduction}

Today, as we witness the dawn of a precision era in observational cosmology, we 
are becoming increasingly aware that our present knowledge and understanding of 
the Galactic foregrounds demand a more thorough underpinning. During the 
course of this meeting we have learned about the results from the latest 
experiments for measuring the anisotropy in the Cosmic Microwave Background 
Radiation (CMBR). Yet, despite their success in ruling out a significant 
contamination of Galactic origin, there are still opened questions regarding the 
spatial and spectral distribution of the Galactic constituents. New insights have 
already improved on this picture as the work of Lazarian on Galactic dust 
properties and of Reynolds on Galactic free-free mapping through H$\alpha$ 
association have been able to tell us. Such an effort also lies at the core of the 
Galactic Emission Mapping (GEM) project in order to reveal a new look of the 
Galactic synchrotron component, capable of overcoming the well-known 
shortcomings of the existing surveys; let alone the important implications for 
the search of spectral distortions in the CMBR.

\section{The GEM Project at \fb}

The GEM project (De Amici et al. 1994; Smoot 1999) was conceived 9 years ago 
at the XXIst IAU General Assembly in Buenos Aires, Argentina, when an 
international team of researchers set forth to develop a portable radiotelescope for 
achieving full-sky coverage between \fa\ and \fe. Since its first deployment in the 
Owens Valley desert near Bishop, CA, (USA) in 1994 the GEM experiment has 
operated from 3 other locations (Villa de Leyva in Colombia, Tenerife in the 
Canary Islands and Cachoeira Paulista in Brazil) and gathered observations at 4 of 
the 5 planned frequencies: \fa\ (Torres et al. 1996), \fb\ (Tello 1997) and \fc\ 
(Torres, internal report, 1997). For the purpose of these proceedings I will 
restrict my attention to the data taken at the Owens Valley and Cachoeira 
Paulista sites at \fb\ which provide a nearly 75\% sky-coverage. 

\begin{figure}[htb]
\begin{center}
\vspace{0cm}
\hspace{0cm}\epsfxsize=5cm \epsfbox{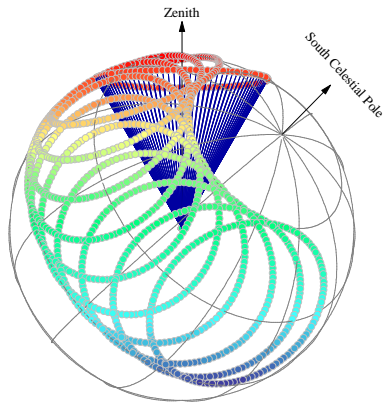}
\vspace{0cm}
\caption{Representation for a site in the Southern Hemisphere of the mapping 
strategy used by the GEM experiment.}
\end{center}
\label{Fig1}
\end{figure}

GEM's mapping strategy uses a 1-rpm rotating dish to scan the sky around the 
zenith in $60\deg$ wide circles, which due to the earth's rotation provide complete 
sky coverage of an equally wide declination band (see Fig.~1). 
The GEM radiotelescope consists of a 5.5-m dish and is double-shielded against 
stray radiation from the ground. We use a rim-halo to intercept the spillover 
sidelobes and a wire mesh fence to further cut down and level out the 
contamination from the ground. By applying a geometric diffraction model to 
this experimental setup (Tello et al. 1999), together with measurements of the 
backfire helical feed pattern and selected field scan observations (Tello et 
al.~2000), we have nailed down the level of ground contamination, which so far 
has demonstrated to be one of the most significant sources of contamination in 
survey experiments. The collimated feed pattern has a HPBW of $5.4\deg$.

The radiometric signal is sampled at nearly half-second intervals by a 
thermally controlled total power receiver within a 90\th{MHz} bandwidth. The 
system temperature is $127.42\pm 0.36$\th{K} and its gain is $61.010\pm 
0.047$\th{K}. We have also performed a detailed study of the receiver systematics 
in order to predict its response to changes in the thermal bath of its components, 
such that the receiver baseline drifts at $0.3591\pm 0.0007$\th{K/$\deg$C} and 
the relative gain change has a thermal susceptability of $0.01030\pm 
0.00001/\deg$C.

\section{Data Analysis and Results}

For some 20\% of the data collected at \fb\ the stability of the receiver guarantees 
a 95\% confidence level of spurious signal rejection due to gain variations on a 
timescale of 150-minute long observing runs. By keeping the Sun and the Moon 
at angles larger than $30\deg$ and $6\deg$, respectively, from axis, a total of 
8.5\% of data had to be purged; while only 1.6\% had to be excised due to RFI 
signals. In addition, the firings from a stable noise source diode at 44.8-s intervals 
were used to correct the received signals. Finally, we applied a de-striping 
algorithm (Tello 1997) to remove low frequency 1/f noise from the remaining 
observations. The technique relies on a sampling of the circular scans into 
networks of 120 half-a-beamwidth wide bins which, when divided into Eastern 
and Western sets, become fiducial baselines to calibrate the Western and Eastern 
scans that intercept them (see Fig.2). These calibrated scans can in turn be used to 
compose further bin networks to the East and West in order to propagate the map 
baseline across an entire declination band.

\begin{figure}[htb]
\begin{center}
\vspace{-.5cm}
\hspace{0cm}\epsfxsize=9.5cm \epsfbox{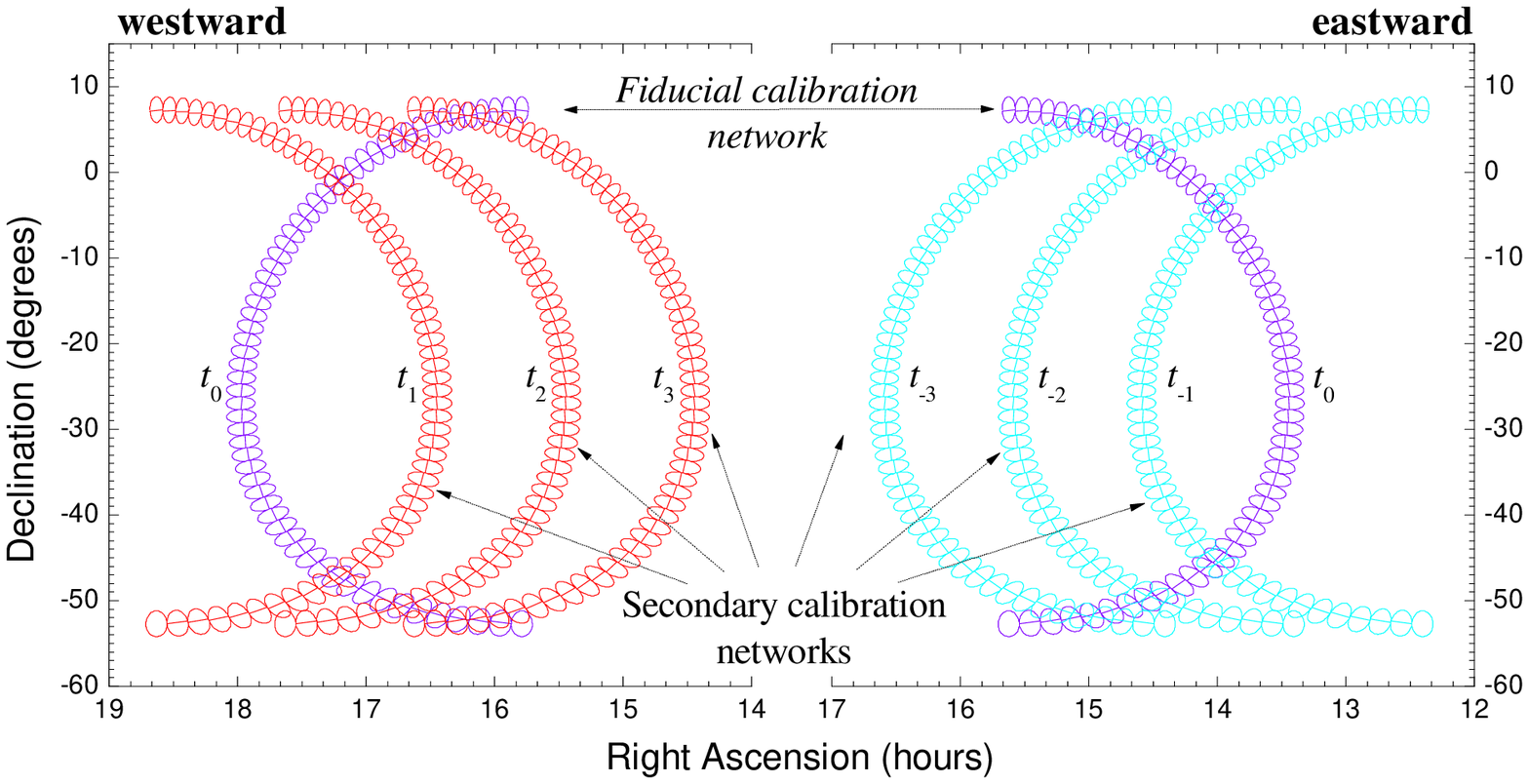}
\vspace{0cm}
\caption{Intersecting networks of 120 bins propagate the map baseline based on 
a fiducial sky region of low temperature contrast.}
\end{center}
\label{Fig2}
\end{figure} 

This technique presumes, however, the absence of stray radiation from the 
Galactic Plane. We have not yet compensated for this source of systematic effect
in the data, which implies in a colder baseline than it should after it has 
propagated from a warm into a cold region of the sky, but the present status of our
work can be best appreciated in the map of Fig.~3. This map combines 63.5 hours
of data from the Bishop site with 212.65 hours from the Brazilian site and has been
set to a pixel resolution of $2.8125\deg$ spanning an antenna temperature range from 
$2.45$\th{K} to $9.65$\th{K} in $0.15$\th{K} steps.

\begin{figure}[htb]
\begin{center}
\vspace{0cm}
\hspace{0cm}\epsfxsize=13cm \epsfbox{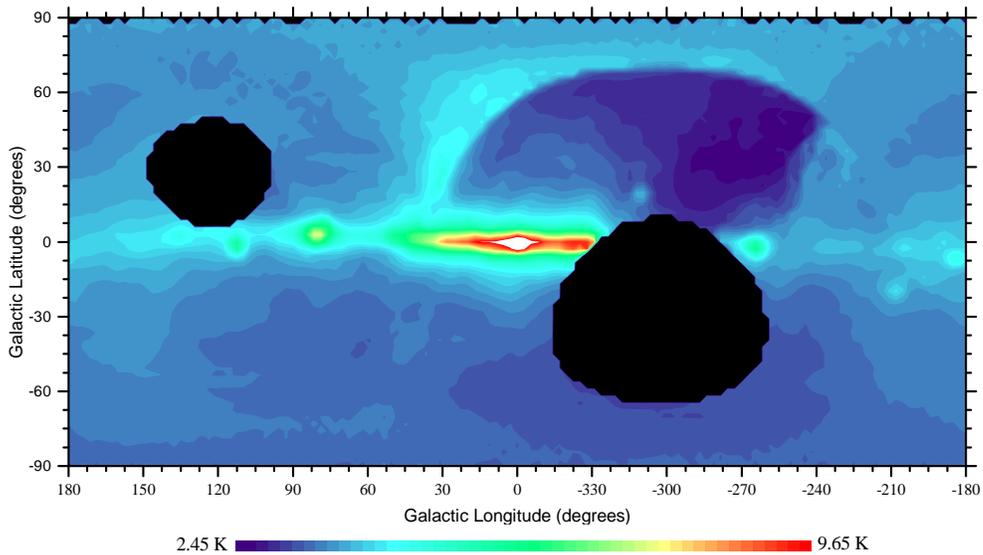}
\vspace{0cm}
\caption{Present status of the GEM data analysis at \fb.}
\end{center}
\label{Fig3}
\end{figure} 

 \acknowledgments
The GEM project, whose principal investigators appear as co-authors in the present
contribution, has been the work of an outstanding team of researchers; mainly M.~Bensadoun, 
M.~Limon, G.~De Amici, J.~Gibson, J.~Yamada, C.~Witebsky, D.~Heine, M.~Leung, A.~Alves, 
L.~Arantes, C.A.~Wuensche, N.~Figueiredo, R.~de Souza, A.P.~da Silva and, last but not 
least, J.~Aymon. 
This work was supported by FAPESP through grant No.~00/05436-1.

\end{document}